\documentclass[aps,twocolumn,pra,superscriptaddress,showpacs,tightenlines]{revtex4-1}
\usepackage{amsmath}
\usepackage{graphicx}
\usepackage{color}
\usepackage{amsfonts}

\begin{document}
\title{Spectrometric reconstruction of mechanical-motional states in optomechanics}
\author{Jie-Qiao Liao}
\affiliation{CEMS, RIKEN, Saitama 351-0198, Japan}
\author{Franco Nori}
\affiliation{CEMS, RIKEN, Saitama 351-0198, Japan}
\affiliation{Department of Physics, The University of Michigan, Ann Arbor, Michigan 48109-1040, USA}
\date{\today}

\begin{abstract}
We propose a spectrometric method to reconstruct the motional states of mechanical modes in optomechanics. This is achieved by detecting the single-photon emission and scattering spectra of the optomechanical cavity. Owing to an optomechanical coupling, the \emph{a priori} phonon-state distributions contribute to the spectral magnitude, and hence we can infer information on the phonon states from the measured spectral data. When the single-photon optomechanical-coupling strength is moderately larger than the mechanical frequency, then our method works well for a wide range of cavity-field decay rates, irrespective of whether or not the system is in the resolved-sideband regime.
\end{abstract}


%42.50.Wk    Mechanical effects of light on material media, microstructures and particles
%03.65.Wj     State reconstruction, quantum tomography
%42.50.Pq     Cavity quantum electrodynamics; micromasers

\maketitle

\section{Introduction}

Quantum states carry complete information of a physical system. To obtain the statistical properties of the system, one needs to know its state vector or density operator
when it is in a pure or mixed state, respectively~\cite{Fano1957}. Quantum state tomography (QST) is a procedure for reconstructing the state of quantum systems by measuring a complete set of observables~\cite{Vogel1989,Leonhardtbook,QSTreviews}.
In the past two decades, many advances have
been made in QST for both continuous-variable~\cite{QSTreviews} and discrete-variable~\cite{Chuang1998,White1999,Steffen2006} states. For example, the QST of a harmonic
oscillator has been analyzed in both quadrature and Fock-state
representations. Moreover, quantum state reconstruction has also been studied on optical~\cite{Smithey1993,Schiller1996,Breitenbach1997,Lvovsky2001,Ourjoumtsev2006} and microwave~\cite{Deleglise2008,Brune2008,Hofheinz2009,Wang2009,Mallet2011,Eichler2011,Shalibo2013} photon fields, as well as
the motional states of matter systems, such as atoms~\cite{Leibfried1996,Kurtsiefer1997,Lutterbach1997}, molecules~\cite{Dunn1995,Waxer1997},  and micro- or nanomechanical resonators~\cite{Rabl2004,Singh2010,Miao2010}.

Generally, it is a difficult task to directly reconstruct the quantum state of a massive mechanical resonator
because one cannot access the mechanical excitation directly. Usually, other auxiliary systems, such as optical modes~\cite{Vanner2011,Vanner2013} or atoms~\cite{Singh2010} are needed to transduce the states of mechanical modes. In this sense, optomechanical systems~\cite{Kippenberg2008,Aspelmeyer2013} can provide a natural platform to perform this task because there is an inherent interface between mechanical and optical modes. Even though people have noticed the means for controlling mechanical motion (e.g., cooling~\cite{Wilson-Rae2007,Marquardt2007,Teufel2011,Chan2011} and quantum-state engineering~\cite{Vitali2007,Wang2013,Tian2013,Mari2009,Liao2011,Law2013,Xu2013,Jing2013}) by designing proper driving fields, the method for monitoring the mechanical motion in the nonlinear quantum optomechanics via the optical means remains mostly unexplored. We note that QST in optomechanics has recently been studied in the short-pulse case~\cite{Vanner2011,Vanner2013} and in optically-levitating-dielectrics systems~\cite{Cirac2011}.

In this paper, we propose a reliable method for reconstructing the mechanical motional state of the moving mirror in cavity optomechanics. Our approach is based on spectrometric measurement of the emission or scattering of a single photon interacting with the mechanical motion. Owing to the optomechanical coupling, the mechanical oscillation will modulate the behavior of the single photon (which will show in its spectrum), and hence one can infer the state information of the mechanical mode from the measured spectral data.

\section{The system}

\begin{figure}[tbp]
\center
\includegraphics[bb=37 676 233 763, width=0.48 \textwidth]{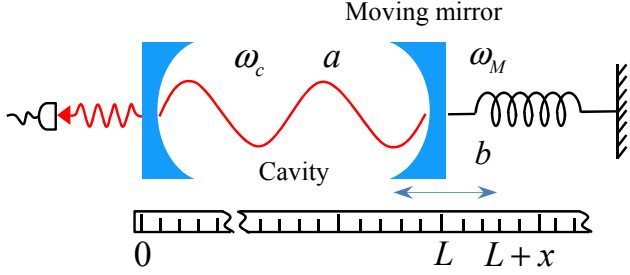}
\caption{(Color online) Schematic diagram of a Fabry-Perot-type
optomechanical cavity formed by a fixed end mirror and a moving end
mirror. The cavity fields couple with the continuous fields outside the cavity through photon-hopping interactions.}\label{setup}
\end{figure}

We start by considering an open optomechanical cavity, which is formed by a fixed end mirror and a moving end mirror, as shown in Fig.~\ref{setup}.
Assuming that the moving mirror is perfect and the fixed mirror is partially transparent, then the cavity fields couple with the continuous fields outside
the cavity through photon-hopping interactions. In a rotating frame with respect to
$H_{0}=\omega_{c}a^{\dagger}a+\omega_{c}\int_{0}^{\infty}c_{k}^{\dagger}c_{k}dk$ ($\hbar=1$), the Hamiltonian of the total system including the cavity and the environment fields is~\cite{Liao2012,Liao2013}
\begin{equation}
H_{s}=H_{\rm{opc}}+\!\int_{0}^{\infty}\!\Delta_{k}c_{k}^{\dagger}c_{k}\;dk+\int_{0}^{\infty}\!\!\xi_{k}(a^{\dagger}c_{k}+c_{k}^{\dagger}a)\;dk,\label{hamiltonian-rot}
\end{equation}
with $H_{\rm{opc}}=\omega_{M} b^{\dagger}b-g_{0}a^{\dagger}a(b^{\dagger}+b)$.
Here, $a$ $(a^{\dagger })$ and $b$ $(b^{\dagger })$ are, respectively, the
annihilation (creation) operators of the optical and mechanical modes, with
respective frequencies $\omega_{c}$ and $\omega_{M}$. The second term in $H_{\rm{opc}}$ describes the radiation-pressure coupling, and $g_{0}$ is the
single-photon optomechanical coupling strength~\cite{Law1995}. The annihilation and creation operators $
c_{k}$ and $c^{\dag}_{k}$ describe the $k$th mode of the outside fields with resonant
frequency $\omega_{k}$. The parameter $\Delta_{k}=\omega_{k}-\omega_c$ is the detuning between the frequencies of the $k$th mode and the cavity mode.
The coupling between the cavity field and the
outside fields is described by the photon-hopping interaction with coupling
strength $\xi_{k}$. Under the framework of the Wigner-Weisskopf theory, the parameter $\xi_{k}$ is related to the photon decay rate by $\gamma_{c}=2\pi\xi_{c}^{2}$,
where $\xi_{c}$ is the strength at the cavity resonance frequency. We note that the Hamiltonian of the total
system in the Schr\"{o}dinger picture can be written as $H=H_{0}+H_{s}$ because of $[H_{0},H_{s}]=0$.

Because the
mechanical decay rate $\gamma_{M}$ is much smaller than the optical decay rate $\gamma_{c}$ ($\gamma_{M}/\gamma_{c}\sim 10^{-3}$ to $10^{-7}$ in typical optomechanical systems~\cite{Aspelmeyer2013}),
in this paper we will merely take the photon decay into account and
neglect the mechanical dissipation. This treatment is justified
because the photon emission and scattering processes are completed in a time scale $1/\gamma_{c}$
during which the mechanical dissipation is negligible.

\section{Single-photon spectra}

In this system, the total photon number operator $T_{p}=a^{\dagger}a+\int_{0}^{\infty}\!c_{k}^{\dagger}c_{k}\,dk$ is a conserved quantity, due to $[T_{p},H]=0$, then we can restrict the system within a subspace with a definite photon number. In the single-photon subspace, a general pure state of the total system can be written as
\begin{eqnarray}
\vert\varphi(t)\rangle&=&\sum_{l=0}^{\infty}A_{l}(t)\vert 1\rangle_{a}\vert\tilde{l}\rangle_{b}\vert\emptyset\rangle\nonumber\\
&&+\sum_{l=0}^{\infty}\int_{0}^{\infty}B_{l,k}(t)\vert 0\rangle_{a}\vert l\rangle_{b}\vert 1_{k}\rangle dk,\label{genstateform}
\end{eqnarray}
where $\vert\tilde{l}\rangle_{b}=\exp[\beta_{0}(b^{\dagger}-b)]|l\rangle_{b}$ are single-photon-displaced phonon number states~\cite{Hu1996} with $\beta_{0}=g_{0}/\omega_{M}$. The states $\vert\tilde{l}\rangle_{b}$ are defined by the eigen-equation
\begin{equation}
[\omega_{M}b^{\dagger}b-g_{0}(b^{\dagger}+b)]|\tilde{l}\rangle_{b}=(l\omega_{M}-\delta)|\tilde{l}\rangle_{b}\label{eigeneq}
\end{equation}
with $\delta =g_{0}^{2}/\omega_{M}$. Here the Hamiltonian operator at the left-hand side of Eq.~(\ref{eigeneq})
is $H_{\rm{opc}}$ when the photon number is limited to one.
In Eq.~(\ref{genstateform}),
$\vert 1_{k}\rangle=c_{k}^{\dagger}\vert\emptyset\rangle$ is the single-photon state of mode $c_{k}$, and $\vert\emptyset\rangle$ is the vacuum state. The first and second components in $\vert\varphi(t)\rangle$ denote the basis states for the single photon in the cavity mode and the $k$th mode outside the cavity, respectively. The variables $A_{l}(t)$ and $B_{l,k}(t)$ are probability amplitudes.

In the long-time limit $t\gg1/\gamma_{c}$, the single photon completely leaks out of the cavity, regardless of its initial state, then $A_{l}(\infty)=0$ and the long-time state becomes
\begin{eqnarray}
\vert\varphi(\infty)\rangle=\sum_{l=0}^{\infty}\int_{0}^{\infty}B_{l,k}(\infty)\vert 0\rangle_{a}\vert l\rangle_{b}\vert 1_{k}\rangle dk.
\end{eqnarray}
The form of $B_{l,k}(\infty)$ depends on the initial state of the system. Here, the single photon could be initially in either the cavity or the outside fields. These two cases correspond to the single-photon emission and scattering processes.

Below, we derive the single-photon emission and scattering spectra when the mechanical resonator is in an arbitrary initial state. This is achieved by first calculating the spectra corresponding to the mirror initially in a Fock state $|n_{0}\rangle_{b}$, and then we obtain the spectra for a general initial state by superposition. When the mirror is initially in the number state $|n_{0}\rangle_{b}$, the initial state of the total system is
\begin{equation}
\vert\varphi_{n_{0}}(0)\rangle=|n_{0}\rangle_{b}|\phi\rangle_{\textrm{photon}}.
\end{equation}
In the single-photon emission case, the single photon is initially in the cavity, and we have $|\phi\rangle_{\textrm{photon}}=|1\rangle_{a}|\emptyset\rangle$. In the scattering case, the cavity is initially in a vacuum and the single photon is in a Lorentzian wave packet in the continuous fields. Then the initial state of the photon is
\begin{equation}
|\phi\rangle_{\textrm{photon}}=|0\rangle_{a}\otimes\sqrt{\frac{\epsilon}{\pi}}
\int_{0}^{\infty}\!\!\frac{1}{(\Delta_{k}-\Delta_{0}+i
\epsilon)}|1_{k}\rangle dk,
\end{equation}
where $\Delta_{0}$ and $\epsilon$ are the wave-packet center and width, respectively. At time $t$, the state of the system can be expressed as Eq.~(\ref{genstateform}),
with the replacements $\vert\varphi(\infty)\rangle\rightarrow\vert\varphi_{n_{0}}(\infty)\rangle$,
$A_{l}(t)\rightarrow A_{n_{0},l}(t)$, and $B_{l,k}(t)\rightarrow B_{n_{0},l,k}(t)$.
Here the subscript $n_{0}$ in state $\vert\varphi_{n_{0}}(t)\rangle$ and probability amplitudes $A_{n_{0},l}(t)$ and $B_{n_{0},l,k}(t)$ refers to the initial state $\vert n_{0}\rangle_{b}$ of the mirror. The expressions of $A_{n_{0},l}(t)$ and $B_{n_{0},l,k}(t)$ for the single-photon emission and scattering cases have been given in Ref.~\cite{Liao2012}.
In the single-photon emission and scattering cases, we can check that the state of the system is normalized, i.e.,
$\langle\varphi_{n_{0}}(t)\vert\varphi_{n_{0}}(t)\rangle=1$.
Since we treat the optomechanical system and the continuous fields outside
the cavity as a whole closed system, the evolution of the total system
is unitary. We should point out that,
in the derivation of the analytical expression of these probability amplitudes, we have made
the Wigner-Weisskopf approximation.

Based on the above discussions, we denote the relation
\begin{equation}
\vert\varphi_{n_{0}}(t)\rangle=U(t)\vert\varphi_{n_{0}}(0)\rangle=U(t)|n_{0}
\rangle_{b}|\phi\rangle_{\textrm{photon}},
\end{equation}
where $U(t)$ is the unitary evolution operator associated with the Hamiltonian $H_{s}$ of the total system.
Therefore, when the mirror is initially in a general density matrix
\begin{equation}
\rho^{(b)}(0)=\sum_{m,n=0}^{\infty}\rho_{m,n}^{(b)}(0)|m\rangle_{b}\,_{b}
\langle n|  \label{generinista}
\end{equation}
with $\rho_{m,n}^{(b)}(0)=\,_{b}\langle
m|\rho^{(b)}(0)|n\rangle_{b}$,
the state of the total system at time $t$ would be
\begin{eqnarray}
\rho(t)&=&U(t)[\rho^{(b)}(0)\otimes|\phi\rangle_{\textrm{photon}}\;_{\textrm{photon}}
\langle \phi| ]U^{\dagger}(t)  \notag \\
&=&\sum_{m,n=0}^{\infty}\rho_{m,n}^{(b)}(0)U(t)|m\rangle _{b}|\phi\rangle_{\textrm{
photon}}\;_{\textrm{photon}} \langle\phi|\,_{b}\langle n|U^{\dagger}(t)  \notag \\
&=&\sum_{m,n=0}^{\infty}\rho_{m,n}^{(b)}(0)\vert\varphi_{m}(t)\rangle\;
\langle\varphi_{n}(t)|.\label{generinistatransT}
\end{eqnarray}
Correspondingly, the long-time state of the total system is
\begin{eqnarray}
\rho(\infty)=\sum_{m,n=0}^{\infty}\rho_{m,n}^{(b)}(0)|\varphi_{m}(\infty)\rangle\langle\varphi_{n}(\infty)|,\label{genelongtsta}
\end{eqnarray}
where $|\varphi_{m(n)}(\infty)\rangle=\sum_{l=0}^{\infty}\int_{0}^{\infty}B_{m(n),l,k}(\infty)|0\rangle_{a}|l\rangle_{b}|1_{k}\rangle dk$. From $\rho(\infty)$ we obtain the single-photon spectra
\begin{equation}
S(\Delta_{k})\equiv\textrm{Tr}[\Pi_{k}\rho(\infty)]=\sum_{m,n=0}^{\infty}\!\!\rho_{m,n}^{(b)}(0)\Lambda_{n,m}(\Delta_{k}),\label{srhoconnect}
\end{equation}
where $\Pi_{k}=|1_{k}\rangle\langle1_{k}|$ is the single-photon projective operator and
\begin{eqnarray}
\Lambda_{n,m}(\Delta_{k})=\sum_{l=0}^{\infty}B_{n,l,k}^{\ast}(\infty)B_{m,l,k}(\infty).
\end{eqnarray}
The relation~(\ref{srhoconnect}) is important in this work and provides the connection between the spectra and the density matrix elements of the mirror. This result motivates us to reconstruct the initial state of the mirror by measuring the spectra of the outgoing photon.

\section{Quantum state reconstruction}

\subsection{Diagonal density-matrix case}

To better see this procedure, we first consider the diagonal density-matrix case, where the density matrix is diagonal in the basis of number states. In this case, the initial state of the mirror is assumed to be
$\rho^{(b)}(0)=\sum_{n=0}^{\infty}P_{n}|n\rangle_{b}\,_{b}\langle n\vert$, with the phonon number distribution $P_{n}$. Then, the single-photon spectra become
$S(\Delta_{k})=\sum_{n=0}^{\infty}P_{n}S_{|n\rangle_{b}}(\Delta_{k})$, where
$S_{|n\rangle_{b}}(\Delta_{k})=\Lambda_{n,n}(\Delta_{k})$
are the spectra corresponding to the component of the Fock state $|n\rangle_{b}$. Once we
know the three parameters: $g_{0}$, $\gamma_{c}$, and $\omega_{M}$, then $S_{|n\rangle_{b}}(\Delta_{k})$ can be obtained.
In realistic reconstructions, we approximately truncate the
Hilbert space of the mirror into the lowest $N$-dimensional
subspace. Here the dimension parameter $N$ should be chosen to be sufficiently large such that the probabilities outside
this subspace are negligible. In this case, the spectra can be approximated by
\begin{equation}
S(\Delta_{k})\approx\sum_{n=0}^{N-1}P_{n}S_{|n\rangle_{b}}(\Delta_{k})\label{mixedconn}.
\end{equation}
Inspired by this relation, we construct
a system of linear equations for the variables $P_{n}$
by choosing $N$ sample points (with
the coordinates $\Delta_{k_{j}}$, $j=1,2,3,...,N$) from the spectra. The compact form of these equations is
\begin{equation}
\mathbf{K}\mathbf{P}=\mathbf{Q},
\end{equation}
where $\mathbf{P}=(P_{0},P_{1},...,P_{N-1})^{T}$. The elements of $\mathbf{K}$ and $\mathbf{Q}$ are defined by
\begin{equation}
\mathbf{K}_{j,j'}=S_{|j'-1\rangle_{b}}(\Delta_{k_{j}}),\hspace{0.5 cm}\mathbf{Q}_{j}=S(\Delta_{k_{j}}),
\end{equation}
for $j,j'=1,2,3,...,N$.
The square matrix $\mathbf{K}$ can be calculated based on the parameters $g_{0}$,
$\gamma_{c}$, and $\omega_{M}$, and the vector $\mathbf{Q}$ can be
measured in experiments. Therefore, if the square matrix $\mathbf{K}$ is full rank, then the phonon number distribution can be obtained as
\begin{equation}
\mathbf{P}=\mathbf{K}^{-1}\mathbf{Q},
\end{equation}
where $\mathbf{K}^{-1}$ is the inverse matrix of $\mathbf{K}$ (see the Appendix for an example).

\begin{figure}[tbp]
\center
\includegraphics[width=0.48 \textwidth]{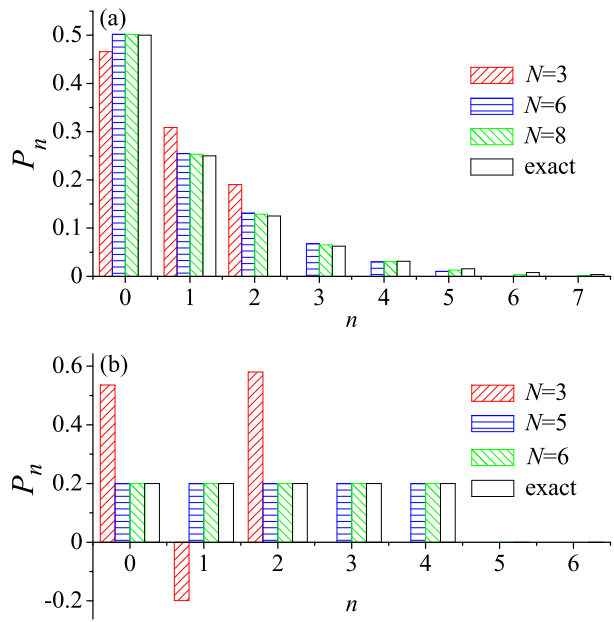}
\caption{(Color online) Reconstructed phonon number distributions $P_{n}$, based on the emission spectrum, for various values of the dimension parameter $N$; the exact distribution is presented for reference. Here the initial state of the mirror is: (a) the thermal state $\rho^{(b)}_{{\rm th}}=\sum_{n=0}^{\infty}\bar{n}^{n}_{\rm{th}}/(\bar{n}_{\rm{th}}+1)^{n+1}|n\rangle_{b\,b}\langle n|$, with $\bar{n}_{\rm{th}}=1$; (b) the maximally-mixed state $\rho^{(b)}_{\rm{mms}}=\frac{1}{\sqrt{n_{s}}}\sum_{l=0}^{n_{s}-1}|l\rangle_{b}\,_{b}\langle l|$ in a subspace of dimension $n_{s}=5$. Other parameters are $\gamma_{c}/\omega_{M}=0.1$ and $g_{0}/\omega_{M}=2$.}\label{fig2}
\end{figure}

One crucial factor in our method is to keep the validity of the truncation approximation in Eq.~(\ref{mixedconn}), namely, we need to choose a proper truncation dimension $N$. A natural question arises: for a unknown initial state of the mirror, how to choose a proper $N$? In a realistic simulation, we need to choose tentative values of $N$ many times by increasing $N$ step-by-step. For an insufficiently large $N$, the reconstructed elements are incorrect, and hence do not converge. We keep on increasing $N$ step-by-step, until the reconstructed data converges. Then this value of $N$ will be good enough to satisfy Eq.~(\ref{mixedconn}). If the $N$ used is larger than the dimension of the exact space, namely the number of nonzero probabilities, then the additional phonon probabilities will be zero in the solution. In this case, $\mathbf{K}$ is still a square matrix.

As examples, below we demonstrate our method by considering two typical mixed states: the thermal state
\begin{equation}
\rho^{(b)}_{\rm{th}}=\sum_{n=0}^{\infty}\frac{\bar{n}^{n}_{\rm{th}}}{(\bar{n}_{\rm{th}}+1)^{n+1}}|n\rangle_{b\,b}\langle n|,
\end{equation}
and the maximally-mixed state
\begin{equation}
\rho^{(b)}_{\rm{mms}}=\frac{1}{n_{s}}\sum_{n=0}^{n_{s}-1}|n\rangle_{b}\,_{b}\langle n|
\end{equation}
in a subspace of dimension $n_{s}$ (with $P_{n}=1/n_{s}$ for $n=0,1,2,...,n_{s}-1$, and $P_{n}=0$ for others). The reconstructed phonon distributions are shown in Fig.~\ref{fig2}, which are compared with the exact phonon distributions. It can be seen from Fig.~\ref{fig2}(a) that
the reconstructed distributions become more stable and eventually converge with increasing $N$. The stable data corresponds to a good truncation approximation in Eq.~(\ref{mixedconn}). The fidelities between the reconstructed and exact  phonon number distributions are $F=0.841,0.980,0.993$ for $N=3,6,8$ in Fig~\ref{fig2}(a).
This truncation effect is more obvious for the state $\rho^{(b)}_{\rm{mms}}$. As shown in Fig.~\ref{fig2}(b), the reconstructed distributions become stable as long as $N\geq n_{s}$. For this state, when $N\geq n_{s}$, the relation~(\ref{mixedconn}) is exact because the Hilbert space is truncated automatically.

\begin{figure}[tbp]
\center
\includegraphics[width=0.48 \textwidth]{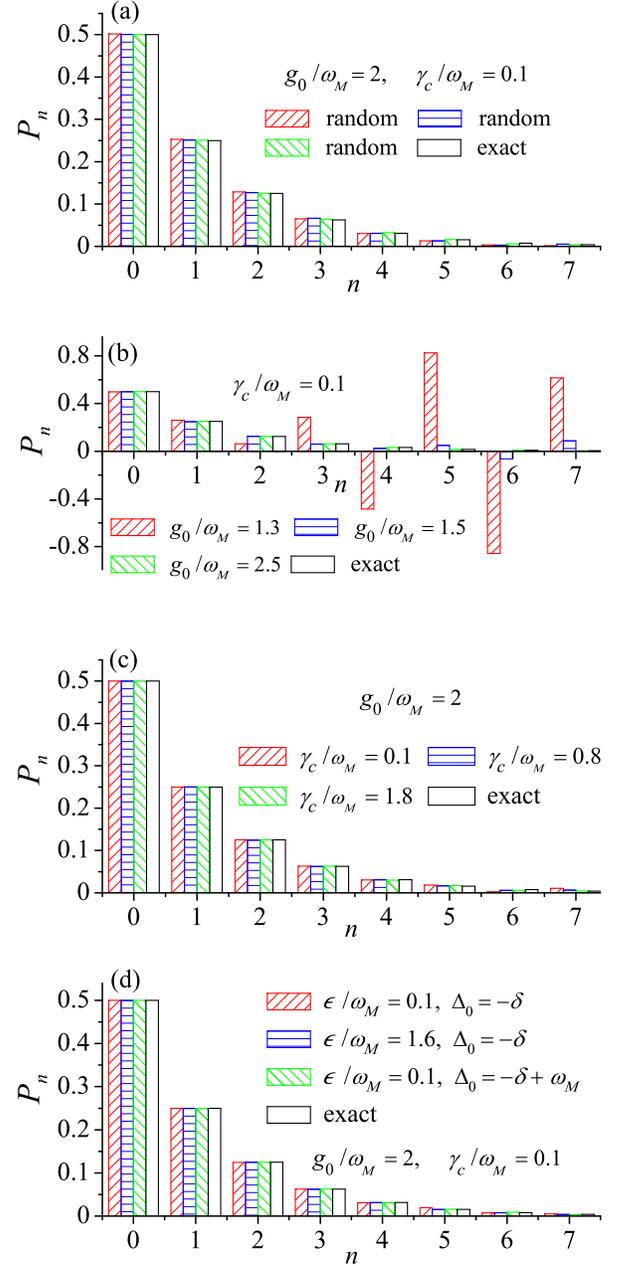}
\caption{(Color online) Reconstructed thermal phonon-number distributions $P_{n}$ based on (a)-(c) emission and (d) scattering spectra in various cases. (a) The sample points are chosen randomly in the region $\Delta_{k_{j}}/\omega_{M}\in[-5,5]$. The reconstructed distributions for various values of (b) the single-photon optomechanical-coupling strength $g_{0}$ or of (c) the cavity-field decay rate $\gamma_{c}$. (d) The distributions are reconstructed based on the scattering spectrum for various values of the parameters $\epsilon$ and $\Delta_{0}$. Here we choose $\bar{n}_{\rm{th}}=1$, and other parameters are shown in the panels.}\label{fig3}
\end{figure}

Equation~(\ref{mixedconn}) indicates that, once the dimension parameter $N$ is large enough to satisfy the truncation approximation,  the reconstructed distributions should be independent of the choice of the sample points. Nevertheless, we should choose the sample points such that the vector $\mathbf{Q}$ can capture the spectral feature as much as possible; mathematically, making sure the matrix $\mathbf{K}$ is full rank. In Fig.~\ref{fig2}, the sample points are located at
\begin{equation}
\Delta_{k_{j}}=-\delta+j\omega_{M}, \hspace{0.5 cm}   j=\lfloor -N/2,...,N/2\rfloor,
\end{equation}
where $\delta=g^{2}_{0}/\omega_{M}$.
These locations correspond to the phonon sideband peaks and can hence mostly capture the spectral feature.
In Fig.~\ref{fig3}(a), we reconstruct the phonon distributions using three groups of random sample points in the region $\Delta_{k_{j}}/\omega_{M}\in[-5,5]$. The results show that the reconstructed results are well consistent with the exact results (with fidelities $F>0.994$), and that our approach is almost independent of the choice of sample points.

Two inherent dimensionless parameters in this system are the scaled single-photon optomechanical-coupling strength $g_{0}/\omega_{M}$ and the scaled cavity-field decay rate $\gamma_{c}/\omega_{M}$~\cite{Ludwig2008,Murch2008,Rabl2011,Nunnenkamp2011}. In order to see how our approach depends on these parameters, we show the reconstructed phonon distributions for various values of the two ratios. As shown in Fig.~\ref{fig3}(b),  we can see that our approach works well for a moderately large coupling strength. Our numerical results show that the optomechanical-coupling strength should be on the scale of $g_{0}/\omega_{M}\sim 2$. A too-weak optomechanical-coupling cannot capture the information regarding the phonon distribution from the spectra, and cannot make sure $\mathbf{M}$ becomes full rank. On the contrary, a large $g_{0}$ will increase the computational difficulties because the involved Hilbert space for phonons is large. In addition, for the cavity-field decay rate $\gamma_{c}$, there is a wide range to choose. Figure~\ref{fig3}(c) proves that our approach works well in a wide parameter range, irrespective of whether or not the system is in the resolved-sideband regime.

Our method can also be realized based on the scattering spectrum. In the scattering case, in addition to the parameters $g_{0}$, $\gamma_{c}$, and $\omega_{M}$, there are two additional controllable parameters: the wave packet centre $\Delta_{0}$ and width $\epsilon$ of the incident photon. In Fig.~\ref{fig3}(d), we plot the reconstructed phonon distributions for various parameters based on the scattering spectrum. We can see that our method works well in both the narrow ($\epsilon/\omega_{M}\ll 1$) and wide ($\epsilon/\omega_{M}> 1$) wave packet cases. In addition, it also works well for a wide range of driving frequencies, which correspond to different phonon sideband resonant transitions~\cite{Liao2012}.

\subsection{General density-matrix case}

We now consider the general density-matrix case, in which the density matrix contains nonzero off-diagonal elements in the number-state representation. By truncating the Hilbert space, the single-photon spectra in Eq.~(\ref{srhoconnect}) can be approximated by
\begin{equation}
S(\Delta_{k})\approx\sum_{m,n=0}^{N-1}\!\!\rho_{m,n}^{(b)}(0)\;\Lambda_{n,m}(\Delta_{k}).\label{pureconn}
\end{equation}
By choosing $N^{2}$ sample points (with locations $\Delta_{k_{j}}$, $j=1,2,3,...,N^2)$ in the spectra, we can construct a system of linear equations of these density matrix elements as
\begin{equation}
\mathbf{MC}=\mathbf{R},
\end{equation}
where the coefficient matrix $\mathbf{M}$ and the column vector $\mathbf{R}$ are defined by
\begin{equation}
\mathbf{M}_{j,j^{\prime}}=\Lambda_{n,m}(\Delta_{k_{j}}),   \hspace{0.5 cm} \mathbf{R}_{j}=S(\Delta_{k_{j}}),
\end{equation}
for $j,j'=1,2,3,...,N^{2}$.
The relationship between the variables $m,n$ and $j'$ is
\begin{equation}
m=\textrm{Floor}[(j'-1)/N], \hspace{0.5 cm}n=(j'-1)-mN,
\end{equation}
where the function $\textrm{Floor}(x)$ gives the greatest integers less than or equal to $x$.
The vector $\mathbf{C}$ is the variable to be determined, its elements
\begin{equation}
\mathbf{C}_{l}=\rho_{m,n}^{(b)}(0), \hspace{0.5 cm}l=1,2,3,...,N^{2},
\end{equation}
are the density matrix elements for the initial state of the moving mirror, where the relationship between the
variables $m$, $n$, and $l$ is $m=\textrm{Floor}[(l-1)/N]$ and $n=(l-1)-mN$.
\begin{figure}[tbp]
\center
\includegraphics[width=0.5 \textwidth]{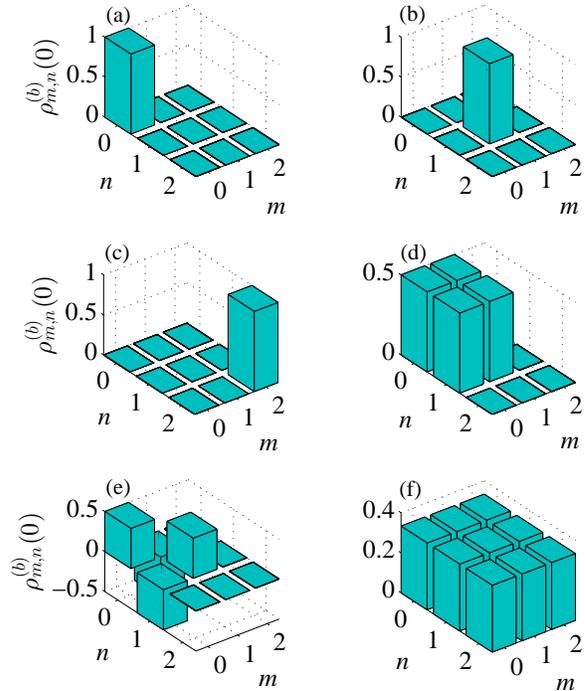}
\caption{(Color online) Reconstructed density matrix elements $\rho_{m,n}^{(b)}(0)$ based on the emission spectrum for various states: (a)-(c) Fock states  $|0\rangle_{b}$, $|1\rangle_{b}$, and $|2\rangle_{b}$, (d), (e) superposed Fock states $(|0\rangle_{b}\pm|1\rangle_{b})/\sqrt{2}$, and (f) $(|0\rangle_{b}+|1\rangle_{b}+|2\rangle_{b})/\sqrt{3}$. Other parameters are $\gamma_{c}/\omega_{M}=0.1$ and $g_{0}/\omega_{M}=2$.}\label{fig4}
\end{figure}

If the square matrix $\mathbf{M}$ is full rank, then
the unique solution of these density matrix elements can be obtained as
\begin{equation}
\mathbf{C}=\mathbf{M}^{-1}\mathbf{R},
\end{equation}
where $\mathbf{M}^{-1}$ is the inverse matrix of $\mathbf{M}$.
The vector $\mathbf{R}$ can be obtained by experimentally detecting the single-photon spectra,
and the coefficient matrix $\mathbf{M}$ can be determined from the expression of $\Lambda_{n,m}(\Delta_{k})$ for special values $\Delta_{k}=\Delta_{k_{j}}$;
then we can reconstruct the initial density matrix of the mechanical mode (see the Appendix for an example).
As additional examples, we consider the density-matrix reconstruction for Fock states and superposed Fock states based on the emission spectrum.
The reconstructed density matrix elements are shown in Fig.~\ref{fig4} for states $|0\rangle_{b}$, $|1\rangle_{b}$, $|2\rangle_{b}$, $(|0\rangle_{b}\pm|1\rangle_{b})/\sqrt{2}$, and  $(|0\rangle_{b}+|1\rangle_{b}+|2\rangle_{b})/\sqrt{3}$.
The data obtained is consistent with the exact result with high fidelities $F>0.999$.

\section{Discussions and conclusions}

There are two important factors in the experimental implementation of our method: (i) The single-photon optomechanical-coupling strength $g_{0}$ should be moderately larger than the resonant frequency $\omega_{M}$ of the mechanical resonator, i.e., $g_{0}/\omega_{M}\sim2$. An optomechanical coupling at this scale can be realized in ultracold atoms. In Ref.~\cite{Murch2008}, a coupling of $g_{0}/\omega_{M}>10$ has been realized. In particular, our method does not require the resolved-sideband condition $g_{0}>\gamma_{c}$, making this proposal experimentally feasible. However, in most optomechanical systems~\cite{Aspelmeyer2013}, the optomechanical-coupling strengths are much smaller than the mechanical frequencies. Recently, two theoretical proposals have been proposed to enhance the optomechanical couplings (the estimated $g_{0}$ is several MHz) in electromechanics~\cite{Sillanpaa2013,Rimberg2013}. Moreover, the ratio $g_{0}/\omega_{M}$ can also be increased by introducing either modulated optomechanical couplings~\cite{Liaomodu}, mechanical normal modes~\cite{Lu2013}, or collective mechanical modes in a transmissive scatter array~\cite{Xuereb}. (ii) How to measure the single-photon emission and scattering spectra is a key step for the realization of this method. In experiments, to profile the pattern of single-photon spectra, the single photon needs to be detected by sweeping the frequency~\cite{Hadfield2009}.

We note that there is no post-selection in our method. This is because we do not need to condition a probability space. In our system, the total photon number is a conserved quantity, and hence we can restrict the system within the single-photon subspace. Namely, a single photon is initially prepared in the cavity (emission) or the continuous field outside the cavity (scattering), after the interaction with the mechanical motion, the single photon will leak out of the cavity due to the photon decay channel. Finally, we measure the spectrum of the single photon, i.e., the reservoir photon occupation distribution. During these processes, there is one and only one photon. Hence, the states of the mechanical motions are completely obtained by measuring the spectrum of the single photon, without post-selection.

To conclude, we have proposed a spectrometric approach for reconstructing the mechanical motional state in optomechanics by detecting single-photon spectra. We considered two different situations: single-photon emission and scattering, which correspond to the cases where the single photon is initially in the cavity field or in a wave packet in the continuous fields outside the cavity. In our considerations, the mechanical dissipation was safely neglected, because the mechanical dissipation is negligible during the time interval for the single-photon emission and scattering to be finished. However, our studies included the photon dissipation by modeling the optical environment as a harmonic-oscillator bath, under the framework of the Wigner-Weisskopf approximation. Our method has a mild constraint in the cavity-field decay rate: it works well in both the sideband-resolved and unresolved regimes. However, the single-photon coupling strength $g_{0}$ should satisfy $g_{0}/\omega_{M}\sim 2$, such that the single-photon spectra can capture the phonon-state information. In the single-photon strong-coupling regime, the non-Gaussian effects are observable. Much recent attention has been paid to explore the non-Gaussian physics in this parameter regime. Nevertheless, quantum state tomography in this regime remains mostly unexplored and here we have proposed a method for
tomography in this regime. Moreover, our method is general and can be potentially realized in various optomechanical systems~\cite{Aspelmeyer2013}. And the idea of spectrometric reconstruction of quantum states can be applied to quadratic optomechanics~\cite{Thompson2008} and cavity-QED~\cite{Raimond2001}.

\begin{acknowledgments}
J.Q.L. thanks Professors C. K. Law, Y. X. Liu, H. Jing, S. Yi, and Dr. X. Y. L\"{u} for valuable discussions. J.Q.L. is supported by the JSPS Foreign Postdoctoral Fellowship No. P12503. F.N. is partially supported by the
RIKEN iTHES Project, MURI Center for Dynamic Magneto-Optics, and a Grant-in-Aid for Scientific Research (S).
\end{acknowledgments}

\begin{appendix}

\section{Examples for quantum state reconstruction based on single-photon emission spectrum\label{Appendix}}

In this Appendix, we present two examples to show how to simulate quantum state reconstruction of a diagonal or a general density matrix in the number-state representation, based on the single-photon emission spectrum.

\subsection{Single-photon emission solution}

In the single-photon emission case, the probability amplitude $B_{n_{0},l,k}(\infty)$ has been given in Ref.~\cite{Liao2012}.
In a realistic simulation, we need to truncate the summation dimension up to a definite value $n_{d}$, namely, up to a phase factor $\exp[-i(l\omega_{M}+\Delta_{k})t]$,
\begin{eqnarray}
B^{n_{d}}_{n_{0},l,k}(\infty)=\sum_{n=0}^{n_{d}-1}\frac{\sqrt{\frac{\gamma_{c}}{2\pi}}
\,_{b}\langle l|\tilde{n}\rangle_{b}\,_{b}\langle\tilde{n}|n_{0}\rangle
_{b}}{[\Delta_{k}+\delta-(n-l)\omega_{M}+i\frac{\gamma_{c}}{2}]},\label{defBndn0lk}
\end{eqnarray}
where we add the superscript $n_{d}$ to mark the summation dimension $n_{d}$.

To describe these Franck-Condon factors $_{b}\langle l|\tilde{n}\rangle_{b}$ and $_{b}\langle \tilde{n}|n_{0}\rangle
_{b}$, we introduce a function
\begin{eqnarray}
&&\mathcal{F}[m,n,\beta]=\;_{b}\langle m|D_{b}(\beta)|n\rangle_{b}\nonumber\\
&=&\sqrt{\frac{(\texttt{min}[m,n])!}{(\texttt{max}[m,n])!}}\;e^{-\beta^{2}/2}(\texttt{sign}[m-n-1/2]\beta)^{\texttt{abs}[m-n]}\nonumber\\
&&\times\texttt{LaguerreL}\left[\texttt{min}[m,n],\texttt{abs}[m-n],\beta^{2}\right],\label{Funcmnbeta}
\end{eqnarray}
where $\texttt{max}[x,y]$ and $\texttt{min}[x,y]$ give the larger and smaller one
between $x$ and $y$, respectively; $\texttt{sign}[x]$ gives $-1$, $0$, or $1$ depending on whether $x$ is
negative, zero, or positive; $\texttt{abs}[x]$ gives the absolute value of $x$, $\texttt{LaguerreL}[n,a,x]$
gives the generalized Laguerre polynomial $L_{n}^{a}(x)$. In terms of Eq.~(\ref{Funcmnbeta}),
we can express these Franck-Condon factors as
\begin{subequations}
\begin{align}
_{b}\langle l|\tilde{n}\rangle_{b}&=\,_{b}\langle l|D_{b}(\beta_{0})|n\rangle_{b}=\mathcal{F}[l,n,\beta_{0}],\\
_{b}\langle \tilde{n}|n_{0}\rangle_{b}&=\,_{b}\langle n|D^{\dagger}_{b}(\beta_{0})|n_{0}\rangle_{b}=\mathcal{F}[n,n_{0},-\beta_{0}].
\end{align}
\end{subequations}
With the emission solution, we can reconstruct the density matrix of the mechanical mode.

\subsection{Diagonal density-matrix case}

In the diagonal density-matrix case, we consider the thermal state $\rho^{(b)}_{{\rm th}}=\sum_{n=0}^{\infty}P_{n}|n\rangle\langle n|$, with $P_{n}=\bar{n}^{n}_{\rm{th}}/(\bar{n}_{\rm{th}}+1)^{n+1}$. The matrix $\mathbf{K}$ and the vector $\mathbf{Q}$ can be determined by
\begin{subequations}
\begin{align}
\mathbf{K}_{j,j'}&=\sum_{l=0}^{n_{d_{1}}-1}|B^{n_{d_{1}}}_{j'-1,l,k_{j}}(\infty)|^{2},\\
\mathbf{Q}_{j}&=\sum_{n,l=0}^{n_{d_{2}}-1}\frac{\bar{n}^{n}_{\rm{th}}}{(\bar{n}_{\rm{th}}+1)^{n+1}}|B^{n_{d_{2}}}_{n,l,k_{j}}(\infty)|^{2},
\end{align}
\end{subequations}
for $j,j'=1,2,...,N$.
The vector $\mathbf{Q}$ is determined by experiments. Hence, in a realistic simulation, we need to choose a very large summation dimension $n_{d_{2}}$, which should be larger than the summation dimension $n_{d_{1}}$ used in obtaining the matrix $\mathbf{K}$. Once we know the three system parameters: $g_{0}$, $\gamma_{c}$, and $\omega_{M}$, so the matrix $\mathbf{K}$ can be obtained.

\begin{widetext}
We randomly generate a group of sample points in the region $\Delta_{k_{j}}/\omega_{M}\in[-5,5]$ (eight sample points,
the truncation dimension $N=8$). The locations of these points are
\begin{equation}
\Delta_{k_{j}}/\omega_{M}=[-0.238903,-4.15271,-2.18112,4.7646,-4.45749,-3.34587,4.92128.-0.361181].
\end{equation}
Using the parameters $n_{d_{1}}=48$, $n_{d_{2}}=60$, $g_{0}/\omega_{M}=2$, $\gamma_{c}/\omega_{M}=0.1$, and $\bar{n}_{\rm{th}}=1$,
we numerically obtain the matrix
\begin{equation}
\mathbf{K}\approx\frac{10^{-2}}{\omega_{M}}\left(
  \begin{array}{cccccccc}
   1.47341 & 1.00608 & 1.11404 & 0.68844 & 1.09823 & 0.88438 & 0.68566 & 0.85648 \\
    3.65773 & 4.03497 & 2.57462 & 2.59686 & 1.30956 & 2.41597 & 3.19158 & 2.85843 \\
    3.41250 & 1.63790 & 1.47325 & 1.71805 & 1.63361 & 1.60156 & 1.29623 & 1.45646 \\
     0.06990 & 0.07098 & 0.40837 & 1.06732 & 1.02115 & 0.63968 & 0.65275 & 0.74980 \\
    0.89358 & 0.83841 & 0.42001 & 0.63426 & 0.46637 & 0.64601 & 0.76208 & 0.63524 \\
    1.26672 & 0.81926 & 0.54281 & 0.61829 & 0.87052 & 0.58272 & 0.45410 & 0.68421 \\
    0.09186 & 0.85753 & 3.79587 & 6.18143 & 4.41536 & 3.26403 & 4.17912 & 4.42602 \\
    0.71383 & 0.72236 & 0.57617 & 0.36789 & 0.74751 & 0.44229 & 0.36831 & 0.56118 \\
  \end{array}
\right),
\end{equation}
and the vector
\begin{equation}
\mathbf{Q}\approx\frac{10^{-2}}{\omega_{M}}(1.23094,3.44540,2.50507,0.22298,0.78367,0.99037,1.37427,0.66994)^{T}.
\end{equation}
By numerically solving the equation $\mathbf{K}\mathbf{P}=\mathbf{Q}$,
we obtain the solution
\begin{equation}
\mathbf{P}\approx(0.50106,0.24944,0.12506,0.06358,0.03105,0.01592,0.00566,0.00766)^{T},
\end{equation}
which has a fidelity $F=0.995$ with the exact phonon number distribution
\begin{equation}
\mathbf{P}_{\textrm{exact}}=(0.5,0.25,0.125,0.0625,0.03125,0.015625,0.0078125,0.00390625)^{T}.
\end{equation}
\end{widetext}

\subsection{General density-matrix case}

In the general density-matrix case, the elements of $\mathbf{M}$ are defined by
\begin{eqnarray}
\mathbf{M}_{j,j^{\prime}}=\sum_{l=0}^{n_{d_{1}}-1}
[B^{n_{d_{1}}}_{n,l,k_{j}}(\infty)]^{\ast}B^{n_{d_{1}}}_{m,l,k_{j}}(\infty),\label{defimatM}
\end{eqnarray}
for $j,j'\in[1,2,...,N^{2}]$, where $m=\textrm{Floor}[(j'-1)/N]$ and $n=(j'-1)-mN$.
The elements of the vector $\mathbf{S}$ are defined by
\begin{eqnarray}
\mathbf{S}_{j}=\sum_{m,n,l=0}^{n_{d_{2}}-1}\rho_{m,n}^{(b)}(0)
[B^{n_{d_{2}}}_{n,l,k_{j}}(\infty)]^{\ast}B^{n_{d_{2}}}_{m,l,k_{j}}(\infty),\label{defimatS}
\end{eqnarray}
for $j\in[1,2,...,N^{2}]$.
In Eqs.~(\ref{defimatM}) and (\ref{defimatS}), the probability amplitudes $B^{n_{d_{1(2)}}}_{m(n),l,k_{j}}(\infty)$ have been given in Eq.~(\ref{defBndn0lk}).

As an example, we assume that the initial state of the mirror is $(|0\rangle_{b}+i|1\rangle_{b}-|2\rangle_{b})/\sqrt{3}$, which has the density matrix
\begin{equation}
\rho^{(b)}_{{\rm exact}}(0)=\frac{1}{3}\left(
                \begin{array}{ccc}
                  1 & -i & -1 \\
                  i & 1 & -i \\
                  -1 & i & 1 \\
                \end{array}
              \right).
\end{equation}
We choose a group of sample points which are located at the phonon sideband peaks, with the locations $\Delta_{k_{j}}=-\delta+n\omega_{M}$ for $n\in[-4,4]$.
Here we use the truncation dimension $N=3$, and then choose nine sample points. Using these parameters $n_{d_{1}}=48$, $n_{d_{2}}=60$, $g_{0}/\omega_{M}=2$, and $\gamma_{c}/\omega_{M}=0.1$, then the locations of these sample points are
\begin{equation}
\Delta_{k_{j}}/\omega_{M}=[-8,-7,-6,-5,-4,-3,-2,-1,0].
\end{equation}
Based on Eqs.~(\ref{defBndn0lk}), (\ref{defimatM}) and (\ref{defimatS}), we can numerically obtain the matrix
\begin{widetext}
\begin{equation}
\mathbf{M}\approx\frac{10^{-1}}{\omega_{M}}\left(
             \begin{array}{ccccccccc}
               2.65078 & 0.72701+0.06619i & 0.31389+0.04904i & \diamond & 2.73543 & 0.92622+0.09507i & \diamond & \diamond & 3.47639 \\
               2.68039 & 0.78019+0.06690i & 0.50595+0.05111i & \diamond & 3.01083 & 1.1303+0.10038i  & \diamond & \diamond & 2.86101 \\
               2.79261 & 0.84290+0.06965i & 0.53053+0.05412i & \diamond & 3.12871 & 1.48178+0.10416i & \diamond & \diamond & 3.40431 \\
               2.98886 & 1.00804+0.07440i & 0.57164+0.02008i & \diamond & 3.30036 & 1.37987+0.02789i & \diamond & \diamond & 3.50278 \\
               3.26298 & 1.15702+0.05181i & 0.57538-0.10663i & \diamond & 3.49535 & 1.26968-0.08778i & \diamond & \diamond & 2.89283 \\
               3.59519 & 1.17639-0.02772i & 0.24166-0.25788i & \diamond & 3.33098 & 0.76202-0.12633i & \diamond & \diamond & 2.43629 \\
               3.86136 & 0.87698-0.13806i & -0.43399-0.29357i & \diamond & 2.78533 & -0.16378-0.10401i & \diamond & \diamond & 1.68726 \\
               3.85851 & 0.08117-0.21613i & -1.30234-0.18395i & \diamond & 1.95965 & -0.72898-0.11875i & \diamond & \diamond & 1.76493 \\
               3.38095 & -1.0129-0.22813i & -1.67703-0.02836i & \diamond & 1.51133 & -0.61621-0.17913i & \diamond & \diamond & 2.29425 \\
             \end{array}
           \right),
\end{equation}
and the vector
\begin{equation}
\mathbf{S}\approx\frac{10^{-1}}{\omega_{M}}\left(
             \begin{array}{ccccccccc}
               2.85246 & 2.62496 & 2.87073 & 2.9511 & 2.80948 & 2.85702 & 2.90594 & 3.17267 & 3.24202 \\
             \end{array}
           \right)^{T}.
\end{equation}
\end{widetext}
In the matrix $\mathbf{M}$, the elements denoted by ``$\diamond$" can be determined by the following rule: In the same row, the elements, with the column pairs $(2\leftrightarrow4)$, $(3\leftrightarrow7)$, and $(6\leftrightarrow8)$, are complex conjugate of each other. We can explain this property based on the relation $\Lambda_{n,m}(\Delta_{k_{j}})=\Lambda^{\ast}_{m,n}(\Delta_{k_{j}})$ in Eq.~(\ref{defimatM}) and the following relations:
\begin{eqnarray}
\mathbf{M}_{j,2}=\Lambda_{1,0}(\Delta_{k_{j}}),\hspace{1 cm}\mathbf{M}_{j,4}=\Lambda_{0,1}(\Delta_{k_{j}});\nonumber\\
\mathbf{M}_{j,3}=\Lambda_{2,0}(\Delta_{k_{j}}),\hspace{1 cm}\mathbf{M}_{j,7}=\Lambda_{0,2}(\Delta_{k_{j}});\nonumber\\
\mathbf{M}_{j,6}=\Lambda_{2,1}(\Delta_{k_{j}}),\hspace{1 cm}\mathbf{M}_{j,8}=\Lambda_{1,2}(\Delta_{k_{j}}).
\end{eqnarray}
In addition, the elements in the first, fifth, and ninth columns are real. This is because $\mathbf{M}_{j,1}=\Lambda_{0,0}(\Delta_{k_{j}})$, $\mathbf{M}_{j,5}=\Lambda_{1,1}(\Delta_{k_{j}})$, $\mathbf{M}_{j,9}=\Lambda_{2,2}(\Delta_{k_{j}})$, and $\Lambda_{n,n}(\Delta_{k_{j}})$ is real.

By numerically solving the equation $\mathbf{M}\mathbf{C}=\mathbf{S}$,
we obtain the expression of $\mathbf{C}$, which can be expressed in the density-matrix form as
\begin{equation}
\rho^{(b)}(0)\approx0.333333 \left(
             \begin{array}{ccc}
            1  & -i & -1 \\
              i & 1 & -i \\
             -1 & i & 1 \\
             \end{array}
           \right),
\end{equation}
where $\rho_{m,n}^{(b)}(0)=\mathbf{C}_{l}$. The relationship between $m,n$ and $l$ is $m=\textrm{Floor}[(l-1)/N]$ and $n=(l-1)-mN$, where $l=1,2,3,...,9$, and $m,n=0,1,2$. The fidelity between the two density matrices $\rho^{(b)}_{{\rm exact}}(0)$ and $\rho^{(b)}(0)$ is almost unity.

\end{appendix}

\end{document}